\documentclass[twocolumn,showpacs,preprintnumbers,amsmath,amssymb]{revtex4}

\usepackage{graphicx}
\usepackage{dcolumn}
\usepackage{bm}

\begin{document}


\title{Non-Classical Response from Quench-Cooled Solid Helium Confined in Porous Gold}

\author{D. Y. Kim}
\affiliation{%
Center for Supersolid \& Quantum Matter Research and Department of Physics, KAIST, Daejeon 305-701, R. O. K.
}%
\author{S. Kwon}
\affiliation{%
Center for Supersolid \& Quantum Matter Research and Department of Physics, KAIST, Daejeon 305-701, R. O. K.
}%
\author{H. Choi}
\affiliation{%
Center for Supersolid \& Quantum Matter Research and Department of Physics, KAIST, Daejeon 305-701, R. O. K.
}%

\author{H. C. Kim}
\affiliation{National Fusion Research Institute (NFRI), Daejeon 305-333, R.O.K.}
\author{E. Kim}%
 \affiliation{%
Center for Supersolid \& Quantum Matter Research and Department of Physics, KAIST, Daejeon 305-701, R. O. K.
}%

\date{\today}

\begin{abstract}
We have investigated the non-classical response of solid $^4$He confined in porous gold set to torsional oscillation. When solid helium is grown rapidly, nearly 7\% of the solid helium appears to be decoupled from the oscillation below about 200 mK. Dissipation appears at temperatures where the decoupling shows maximum variation. In contrast, the decoupling is substantially reduced in slowly grown solid helium. The dynamic response of solid helium was also studied by imposing a sudden increase in the amplitude of oscillation. Extended relaxation in the resonant period shift, suggesting the emergence of the pinning of low energy excitations, was observed below the onset temperature of the non-classical response. The motion of a dislocation or a glassy solid is restricted in the entangled narrow pores and is not likely responsible for the period shift and long relaxation.
\end{abstract}


\maketitle

The state of matter characterized by the coexistence of crystallinity and superfluidity is termed the supersolid state of matter. The first possible evidence of a supersolid helium phase was observed by torsional oscillator (TO) measurements and replicated by other groups \cite{vycorto,bulkto,reppy1,kojima1,shirahama1,kubota2,davis}.  In an ideal TO containing a rigid solid the resonant period follows the rotational inertia of the torsion cell. Kim and Chan observed the reduction in the resonant period and interpreted it as the appearance of non-classical rotational inertia (NCRI). On the other hand, a non-superfluid mechanism possibly induces a reduction in the resonant period if the rigidity of the solid is temperature-dependent \cite{Nussinov,dorsey2}. For instance, the response of the dislocation network to an oscillatory stress field can produce dissipation and a reduction of the resonant period in TO. This scenario is substantiated by recent shear modulus measurements that reveal striking stiffening with qualitatively similar dependence on the temperature, frequency, oscillating velocity, and $^3$He concentration to those of NCRI. The increase in the shear modulus can be well understood by the pinning of the dislocation network with $^3$He impurities.  Finite element method studies and visco-elastic analysis, however, show that less than 10\% of the entire period shift can be attributed to the stiffening effect \cite{clark2,dorsey2}.  In the case of hcp solid $^3$He, no NCRI was found despite a similar increase in the shear modulus, indicating that NCRI is related to a Bose solid \cite{West1}.

\begin{figure}[b]
\includegraphics[width=0.85\columnwidth]{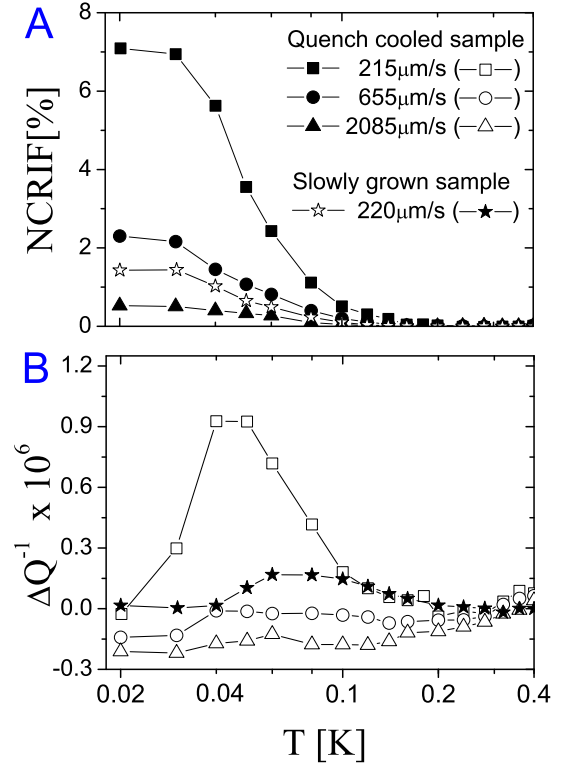}
\caption{\label{fig:epsart} Non-classical rotational inertia fraction (A) and dissipation (B) of the quench-cooled solid $^4$He in porous gold. The symbol with the open star (solid star) represents the NCRI (dissipation) of the slow-cooled sample.}
\end{figure}

There has been considerable theoretical effort to identify the origin of the non-classical response of solid helium \cite{prokofev3,balibar2}. There is a general consensus that the ground state of solid helium is commensurate and superfluidity does not coexist with a commensurate solid \cite{ceperley1,prokofev1,reatto2}. Accordingly, the role of disorders such as vacancies, impurities, and dislocations in solid helium has been investigated \cite{boninsegni4,boninsegni3,pollet1}. Recent extended TO measurements also reported various complicated features likely related to disorders such as large variations in the magnitude of NCRI, confinement/annealing effects, and $^3$He effects. Disorders in helium crystals clearly play an important role, but how the disorders facilitate NCRI remains unclear.

Anderson proposed an alternative explanation in which a thermally excited fluctuating vortex fluid state existing as the lowest excitation in solid helium can exhibit a similar dynamic response in TO above $T_c$ \cite{anderson1}. The flow of quantized vortex tangles induces dissipation and a resonant period drop when the flow rate coincides with the resonant frequency of TO. Anderson's model explains several characteristic aspects qualitatively, but few experiments have been carried out to investigate the vortex fluid state \cite{clark2, kojima2, kubota2, kubota3,VortexTO}.

The present study investigated highly disordered solid helium enclosed in porous gold(PG) where a high density of disorders is expected due to structural confinement. Heavy disorders can also be introduced by a quench-cooling technique during solidification. Thus, NCRI in a sample with structurally induced disorders can be compared to a case with additional quenched disorders. The dynamic response of TO containing a highly disordered solid was also studied to understand the function of low temperature excitations. This study may provide a key to understanding the subtle issues in NCRI, as the complicated porous structure of PG prohibits any elastic motion of line defects larger than a characteristic pore diameter.

The resonant frequency of TO is nearly 948 Hz and the mechanical $Q$ factor measured by a conventional ring-down time constant is $2 \times 10^5$. The physical dimension of the porous gold disk is 10 mm in diameter and 0.6 mm in height. The disk has a surface area of 0.66 m$^2$ and porosity of 67\% according to oxygen vapor isotherm measurements. The pore diameter is about 180 nm assuming monodisperse cylindrical pores.  The PG was covered by a thin layer of Stycast 2850FT and was inserted into a BeCu torsion cell. The empty space was deliberately removed by filling this space with Stycast 2850FT. We found that the 'bulk' open volume in the torsion cell was negligible from the fact that no apparent bulk superfluid transition appeared at 2.176K.  The pressure in the torsion cell was monitored using a resistive strain gauge that measures the deflection of the torsion cell due to high pressure with about a resolution of approximately 0.5 bar. The solid helium sample was grown by the blocked capillary method, and a sudden increase in the TO amplitude marked the completion of solidification.  It was possible to finish the entire freezing process in few seconds via quench-cooling. During the solidification, a minute amount of latent heat (0.01 $\mu$J) was generated due to the small open volume (0.032 cc) to helium and was drained efficiently via the complicated network of gold strands.

We cooled the sample with finite drive voltages that were set at 400 mK. The samples sat at the minimum temperature for about three hours before a scan was started. The resonant period and amplitude data were measured during the warming scan. Measurements were made with various drive voltages inducing different amplitudes (and speeds) of oscillation.

Fig. 1 shows the NCRI fraction (NCRIF) and dissipation of the TO enclosing solid $^4$He sample at 48 bar as a function of temperature. Below 0.2 K the resonant period reveals the onset of NCRI that shows qualitatively identical temperature and velocity dependence to those of previous measurements. The large NCRIF of about 7\% was observed at an oscillating speed of about 200 $\mu$m/s. It was noted that the small rotational inertia and the high torsion constant were likely connected to a limit on the minimum oscillation speed. The characteristic saturation of NCRIF at a low speed limit was not seen down to 200 $\mu$m/s, indicating that the NCRIF at a low speed limit is probably higher than 7\%. For each set of a specific rim speed $v_{max}$, a broad minimum in the amplitude of the oscillation, a signature of dissipation, is detected at the temperature where the period shift shows the maximum variation. A large NCRIF is also reported in highly disordered solids grown by quench-cooling methods in thin annular channels \cite{reppy1}.

\begin{figure}
\includegraphics[width=0.85\columnwidth]{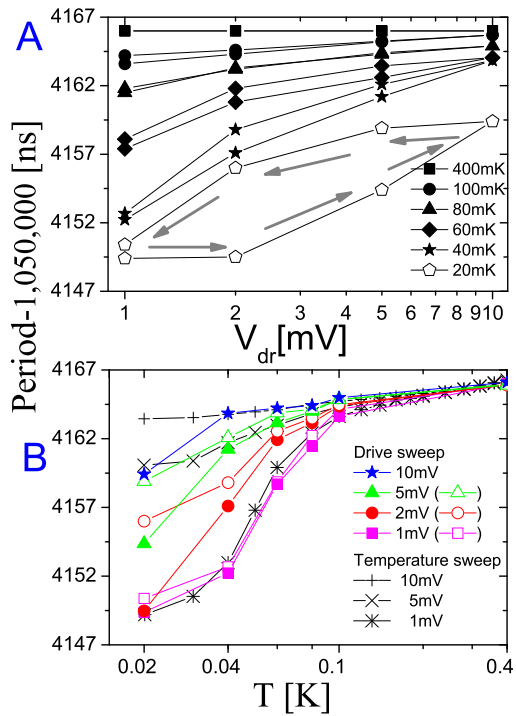}
\caption{\label{fig:epsart} (A) The change of the resonant period during the drive sweep at various temperatures (B) The resonant period in drive sweeps plotted as a function of temperature. (see text) Solid (open) symbols denote final drives in the drive up (down) scans.}
\end{figure}

Solid helium was subsequently grown rather slowly so that the dwelling time on the melting curve was about 90 mins. The final pressure of the sample was tuned to be roughly equal to the pressure of the highly disordered sample. The onset was quite similar to the quenched sample but a smaller magnitude of NCRI (about 1.4\%) with broader dissipation was observed as shown in Fig.1. A small NCRIF has also been found in even more severely constrained solid samples \cite{vycorto, shirahama2}, which supports the explanation that a large NCRI primarily originates from the quench-cooled frozen disorders \cite{West2}. The strong confinements may play a role in stabilizing the frozen disorders. In addition, the results may provide a key to understanding why the thermal histories of different growth methods result in a large variation of the magnitude of NCRI.

NCRI in porous media cannot be understood in terms of a stiffening of solid helium induced by the impurity pinning of the dislocation network. Given that the length scale of a free vibrating dislocation cannot exceed the pore diameter, the characteristic dislocation length in PG can be expected to be much shorter than that in bulk helium. Thus, the increase in the shear modulus due to pinning dislocation networks in PG is expected at substantially lower temperatures (or a higher $^3$He impurity concentration) compared to the bulk value.

Computational studies suggested that the core of screw dislocation lines can be a superfluid \cite{boninsegni2, pollet2,corboz}. Incorporating this suggestion, a scenario was suggested to explain the experimental findings in terms of dislocation induced superfluidity. The pinning of the dislocations can be induced by a connection of neighboring dislocations that facilitates the appearance of NCRI. $^3$He impurities play a key role in stabilizing disorder and promoting superfluidity in this model. The large magnitude of NCRI is, however, not clearly explained in this model.

We also studied the relaxation dynamics of the resonant period when the driving voltage changes discretely at 20, 40, 60, 100, and 400 mK. A solid sample was first cooled to target temperatures with a 1 mV drive. The driving voltage was then increased in sequential steps ($1\rightarrow 2 \rightarrow 5 \rightarrow 10$ mV) and then decreased in the reverse order with the same steps to 1 mV, as shown in Fig. 2-A. Below the onset of NCRI, extended relaxation in the TO period appears in the drive-up scans, while fast relaxation occurs in drive-down scans. The discrepancy in the relaxation leads to hysteretic behavior during a drive sweep at low temperatures. The hysteresis is more pronounced below 60 mK, as the relaxation is lengthened progressively with decreasing temperature. The collection of the drive sweep data is shown in Fig. 2-B; the data points of the drive-up or drive-down scan with the same drive are connected correspondingly for easy comparison to typical temperature sweeps. The reconstructed NCRIF of drive-down scans collapse onto the conventional temperature sweep with the corresponding drive. In contrast, the drive-up scans show discrepancies in their temperature sweeps, reflecting the long relaxation. The apparent drive hysteresis was reported with robust NCRI upon increasing drive, while NCRI was suppressed when the drive decreased \cite{kojima1,clark2}. This hysteresis can be explained by the severe pinning of vortices or by the dislocation network. The severe pinning at low temperatures was only observed at the minimum TO drive in the current experiment.

\begin{figure}
\includegraphics[width=0.85\columnwidth]{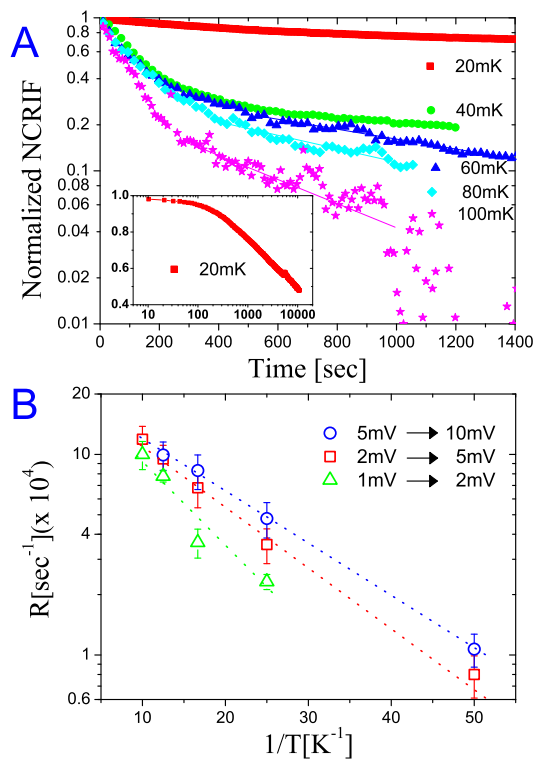}
\caption{\label{fig:epsart} (A)The normalized NCRI when the drive voltage increases from 2 mV to 5 mV is plotted as a function of the elapsed time. Clear deviation from the mechanical ring down is found in the intermediate time scale.  The inset shows logarithmic time evolution (20 mK) at a long time scale. (B) The relaxation in an intermediate time scale as a function of the inverse temperature for various drive voltages. The activation energy is stress-dependent, 98.9 mK for 1 mV $\rightarrow$ 2 mV, 66.7 mK for 2 mV $\rightarrow$ 5 mV, and 60 mK for 5 mV$\rightarrow$ 10 mV.}
\end{figure}

Figure 3-A shows time evolution data of the resonant period change during the drive-up scans. The change of the TO period is normalized by its saturation value to highlight the time-dependent relaxation. For a very short time scale, the relaxation of the TO period shows a simple exponential form that is solely attributed to the high mechanical $Q$ factor regardless of the drive-up or drive-down scans. An exponential evolution with a time constant larger than the mechanical relaxation of the TO arises in an intermediate time scale. The relaxation is further extended as the temperature decreases.  Finally, logarithmic relaxation appears in a long time scale at low temperatures, as depicted in the inset of Fig 3-A. The origin of logarithmic time evolution is not clearly understood, but it has been found in various relaxation phenomena such as dislocation creep \cite{dislocationcreep}, magnetization in spin glass \cite{spinglass}, flux creep in hard superconductors \cite{anderson0}, and magnetization in high Tc superconductors \cite{yeshurun}.

Logarithmic relaxation was observed where disorders function as energy barriers and pin down thermal excitations \cite{anderson0, yeshurun}. The pinned excitations can overcome the barriers with thermal activation assisted by the driving force such as Magnus force and Lorentz force in this framework. The difference between the average pinning potential barrier and the driving kinetic energy ($\Delta U$) gives an empirical unpinning probability that is proportional to exp$(-\Delta U/k_{B} T$). During the drive-up scan, weakly trapped excitations can be released first, exhibiting additional dissipation and a period increase. This process emerges with an exponential relaxation in which a temperature-dependent characteristic time constant (\textit{i.e.} the relaxation rate) is connected to the unpinning probability. As shown in  Fig. 3-B the extended relaxation rate of the unpinning process is proportional to the inverse temperature. The relaxation rate shows a weak dependence on a stress field; higher drive agitation leads to faster relaxation as expected. Once most of the weakly localized excitations are removed the unpinning probability decreases substantially. In this limit, the unpinning of strongly pinned excitations results in logarithmic time dependence at low temperatures, which is consistent with the observed relaxation at the long time scale. More systematic studies on the relaxation behavior have been performed \cite{VortexTO}.

Similar long relaxation of the TO amplitude was observed by a Rutgers group in bulk solid helium with no discernable relaxation in the period \cite{kojima2}. The authors held that the relaxation dynamics was hardly reconciled with the condensation (or evaporation) of $^3$He in dislocation networks and with the quantum mechanical tunneling of dislocations, since the expected relaxation time was inconsistent with the observation. A possible connection of this logarithmic decay to thermally activated vortices was suggested. The extended relaxation can be associated with a new dissipation mechanism that is necessary for re-adjusting the number of vortices to newly introduced drive levels.

Recently a large NCRIF (about 4\%) was found by a Cornell group in TO with a 100 $\mu$m annular gap \cite{davis}. When the temperature of the TO cell increased from the base temperature, ultra-slow relaxation appears below the onset of NCRI. The most surprising observation was that the dissipation peak temperature shifted to lower temperatures progressively as the waiting time increased. The temperatures at which the period showed maximum variation also moved to lower temperatures. This shift is qualitatively similar to the relaxation dynamics of a glassy solid, which is a possible non-superfluid explanation of the period reduction \cite{boninsegni4, balatsky}. The magnitude of the period shift is, however, an order of magnitude larger than that expected in a glassy solid. Thus, NCRI with glass behavior was interpreted as a superglass behavior. The motion of the liquid and/or solid confined in PG is, however, limited by the complicated multiply connected tortuous geometry. Therefore, it is quite unusual to expect any glassy solid-induced dissipation in narrow pores other than superfluidity.

In summary, we found that the magnitude of the non-classical response of solid helium confined in PG to torsional oscillation is strongly dependent on the cooling history of the sample. Substantial enhancement of NCRI in quench-cooled sample and the absence of the confinement effect in porous glasses suggest that the cooling history plays a more significant role in producing large NCRI compared to that by geometrical confinement. We also observed the extended relaxation behaviors of the resonant period in a severely constrained solid helium. The long relaxation time is not likely caused by the motion of an elastic solid but possibly connected to the temperature- and velocity-dependent pinning of low energy excitations at low temperatures.

\begin{acknowledgments}
We acknowledge the useful discussions with M. H. W. Chan. We also wish to acknowledge financial support through Creative Research Initiatives (Center for Supersolid and Quantum Matter Research) of MEST/KOSEF.
\end{acknowledgments}

\end{document}